\documentclass{article}
\usepackage{booktabs}   
\usepackage{multirow}   
\usepackage{makecell}   
\usepackage{ijcai26}
\usepackage{array}
\usepackage{times}
\usepackage{soul}
\usepackage{url}
\usepackage[hidelinks]{hyperref}
\usepackage[utf8]{inputenc}
\usepackage[small]{caption}
\usepackage{graphicx}
\usepackage{amsmath}
\usepackage{amsthm}
\usepackage{booktabs}
\usepackage{algorithm}
\usepackage{algorithmic}
\usepackage[switch]{lineno}
\usepackage{xcolor}
\usepackage{booktabs}

\usepackage{listings}
\usepackage{xcolor}
\usepackage[T1]{fontenc}
\usepackage{lmodern}
\usepackage{multirow}

\newcommand{\tool}{ParPal}

\title{ParPal: A Human-Centred AI System for Multi-Actor Planning and Collaboration in Family Learning}

\author{
Si~Chen,
Jingyi~Xie,
Yao~Li,
Ya-Fang~Lin,
He~Zhang,
Ge~Wang,
Gaojian~Huang,
Rui~Yu,
Ronald~Anthony~Metoyer,
Ting~Hua,
Nitesh~Chawla
}

\begin{document}

\maketitle

\begin{abstract} Family learning takes place in everyday routines where children and caregivers read, practice, and develop new skills together. Despite growing interest in AI tutors, most existing systems are designed for single learners or classroom settings and do not address the distributed planning, coordination, and execution demands of learning at home. This paper introduces ParPal, a human-centred, LLM-powered system that supports multi-actor family learning by decomposing learning goals into actionable subtasks, allocating them across caregivers under realistic availability and expertise constraints, and providing caregiver-in-the-loop tutoring support with visibility into individual and collective contributions. Through expert evaluation of generated weekly learning plans and a one-week field deployment with 11 families, we identify systematic failure modes in current LLM-based planning, including misalignment with role expertise, unnecessary or costly collaboration, missing pedagogical learning trajectories, and physically or temporally infeasible tasks. While ParPal improves coordination clarity and recognition of caregiving effort, these findings expose fundamental limitations in how current LLMs operationalize pedagogical knowledge, reason about collaboration, and account for real-world, embodied constraints. We discuss implications for human-centred AI design and AI methodology, positioning multi-actor family learning as a critical testbed for advancing planning, adaptation, and pedagogical structure in next-generation AI systems.
\end{abstract}

\section{Introductions}

\begin{figure*}[t]
  \centering
  \includegraphics[scale=0.51]{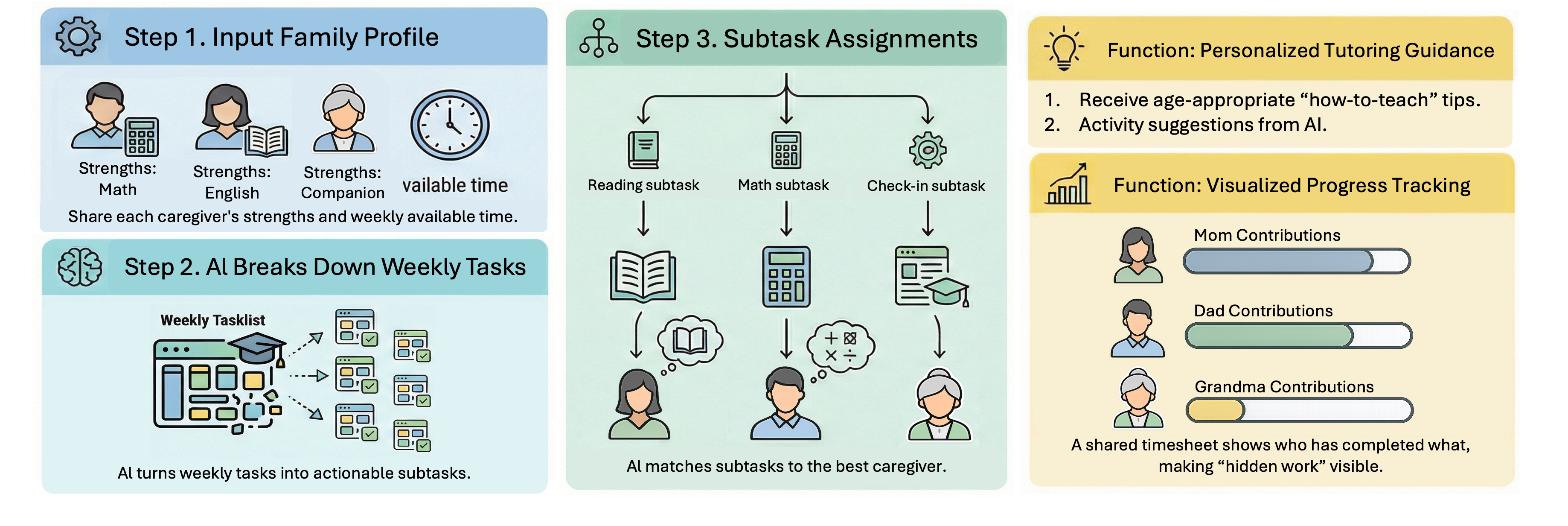}
  \caption{ParPal System Overview}
  \label{Overview}
\end{figure*}

Family learning encompasses the everyday ways children and caregivers engage in learning together--from reading stories and supporting homework to practicing skills and exploring cultural traditions. Unlike classroom learning, family learning unfolds in naturalistic, everyday settings and is embedded in routines, relationships, and embodied interaction. It involves not only the transfer of knowledge but also coordination of responsibilities, negotiation of values, and recognition of contributions \cite{falk2000learning,rogoff2003cultural}. Educational research emphasizes the importance of parental involvement, particularly “learning at home,” where caregivers structure learning processes, model persistence, and connect academic activities to broader life experiences \cite{epstein2018school}. Family learning is therefore shaped by relational, emotional, and organizational practices that reflect human cognition and social interaction, rather than content delivery alone \cite{moll1992funds,hochschild1983managedheart}.

Prior research has examined how families use digital tools--such as calendars, to-do lists, and communication platforms--to support coordination and collaboration in everyday life \cite{neustaedter2009calendar,davidoff2010routine}. In learning contexts, collaboration among caregivers can enrich children's experiences by bringing diverse expertise, sustaining motivation, and strengthening family bonds \cite{falk2000learning,kim2025bridging}. At the same time, family learning presents persistent challenges, including uneven divisions of labor, limited coordination, unrecognized invisible work, and tensions arising from differing educational values or approaches \cite{jo2020parentingstress,lin2024coparenting}. These challenges highlight the multi-faceted nature of human intelligence and caregiving practices in real-world problem solving.

Recently, large language models (LLMs) and AI tutors have been promoted as personalized educational assistants that provide instant answers, practice problems, and scaffolding strategies. Systems such as Khanmigo \cite{khanmigo2025} and Duolingo's AI tutor \cite{duolingo2025} position AI as a one-on-one assistant guiding individual learners. However, most existing systems are designed around a single user, overlooking the distributed, collaborative, and relational characteristics of family learning. Framing homework as “learning at home” and shared responsibility reveals a gap in current AI design: limited support for collaborative assistance, coordination among multiple caregivers, and visibility into human contributions.

These challenges are particularly salient in contexts where families invest substantial time and effort in at-home learning and caregiving. In such settings, responsibilities are often distributed across multiple family members, creating opportunities for intergenerational learning while amplifying issues of workload imbalance, coordination, and reliance on external support. AI tutoring tools that automatically solve problems, generate practice questions, or recommend study pathways are increasingly embedded in everyday routines \cite{zuoyebang2025,quarkgaokao2024}, raising questions about how AI systems shape human behaviour, expectations, and trust in learning support.
Through a formative empirical study with families using AI for homework tutoring, cultural literacy, and skill-building, we identify challenges of limited coordination, imbalanced responsibilities, and parental gatekeeping. Building on these findings, we design ParPal, an AI-augmented system that distributes learning tasks across caregivers, provides individualized support, and surfaces hidden contributions. A one-week field study with 11 families demonstrates how such a system can ease caregiving burdens while fostering recognition, shared accountability, and richer family learning experiences. Our contributions include (1) the design and deployment of ParPal as a human-centred AI system for collaborative assistance in everyday learning, (2) reframing “learning at home” as a collaborative, intergenerational practice grounded in human behaviour and cognition \cite{epstein2018school,moll1992funds}, and (3) demonstrating how AI systems can surface invisible caregiving work, rebalance responsibilities, and calibrate trust between human and AI roles \cite{hochschild1983managedheart}. Together, our findings surface both opportunities and challenges for AI in education, highlighting how human-centred AI design can support collaboration while also revealing tensions around coordination, responsibility, and reliance on AI in real-world learning contexts.

\section{System Design}

\begin{figure*}[t]
  \centering
  \includegraphics[scale=0.19]{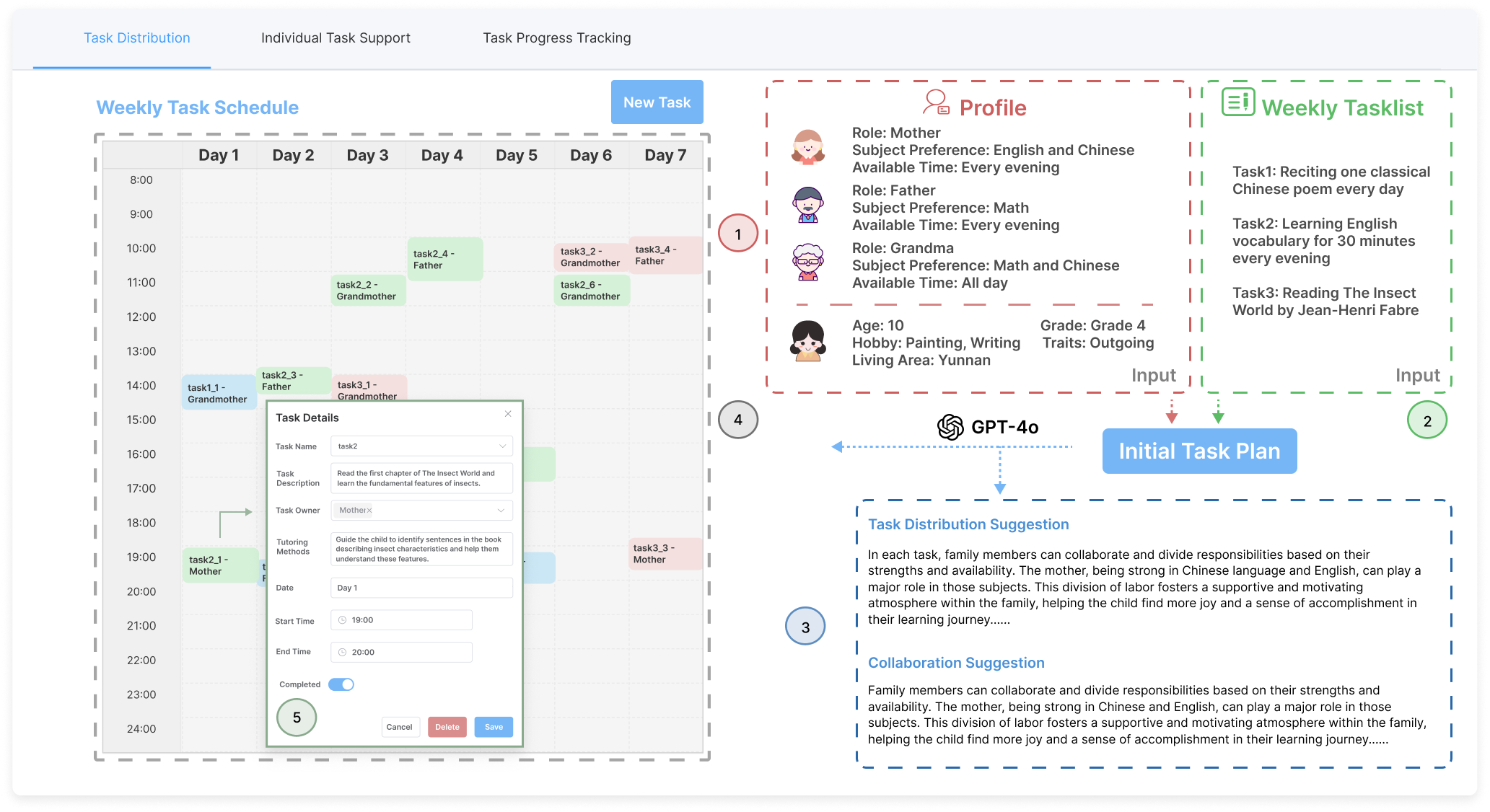}
  \caption{Task Distribution Panel to support task decomposition and allocation.}
  \label{Figure2}
\end{figure*}

\subsection{Formative Study}
We conducted a formative study to understand families' daily family learning practices, their use of AI tools, and expectations for AI-empowered learning tools. We recruited 12 adult caregivers from 7 families through prior contact and snowball sampling. Participants had children in elementary or junior high school, tutoring experience, and prior AI use for family learning. When tutoring was shared, all available caregivers were invited. Participation was voluntary and uncompensated. We conducted one-on-one semi-structured interviews on family learning practices, collaboration, AI use, and expectations. Interviews lasted approximately 60 minutes and were audio-recorded with consent. Two authors conducted inductive thematic analysis \cite{braun2006using}, yielding four key insights (KI).

\label{formative_insights}

\paragraph{\textbf{KI1: Multiple caregivers with subject- or availability-based tutoring.}}
Five of seven families reported involvement of multiple caregivers (commonly parents and grandmothers). Tutoring followed either a subject-based division of labor or an availability-based approach, particularly for younger children. Despite multiple caregivers' involvement, coordination was minimal. Tutoring largely consisted of parallel contributions, with little information sharing; messaging tools were often ineffective due to information overload.

\paragraph{\textbf{KI2: Imbalanced division of labor.}}
Availability-based tutoring often led to workload imbalance, with one caregiver acting as the primary tutor. Contributions were rarely documented, resulting in undervaluation of invisible tutoring work.

\paragraph{\textbf{KI3: Parental gatekeeping of children's AI use.}}
All participants expressed concern about children's over-reliance on AI. Parents therefore restricted independent AI use and guided how AI was incorporated into learning, using AI to enrich content, learn before teaching, or review completed homework.

\subsection{Tool Details}

Based on the formative study, we derived four design strategies to guide the system design: \textbf{DS1} task distribution,  \textbf{DS2} progress visualization, and \textbf{DS3} tutoring support. Overview of the system is shown in Figure \ref{Overview}. \textit{ParPal} (Parenting Partner for Everyday Learning) is an AI-empowered system designed to support collaborative family learning by distributing tasks, coordinating caregiving work, and providing task-specific tutoring assistance. ParPal is implemented as a mobile-responsive web (can be used both on a phone and an iPad) application and uses the GPT-4o API to decompose weekly learning plans into actionable subtasks, allocate and schedule these subtasks across caregivers based on availability and expertise, and generate task-specific tutoring guidance.

ParPal consists of two core components: a \textit{Task Distribution Panel} that generates and visualizes a shared weekly task plan across caregivers, and an \textit{Individual Task Support Panel} that provides AI-assisted guidance for completing assigned subtasks. Details explained below:

\subsubsection{Task Distribution Panel (Figure~\ref{Figure2})}

The Task Distribution Panel supports task decomposition, allocation, and progress tracking for weekly family learning.

\paragraph{Personalized task decomposition and allocation}

Caregivers provide profile information (e.g., subject preferences and available tutoring time) and a weekly task list describing expected learning activities. To achieve this, a \textit{multi-step prompting process} is designed to decompose, allocate subtasks, and support individualized tasks by incorporating profile information \textcircled{1}  and weekly task list \textcircled{2} into prompts. Once all profiles are configured, users can proceed to click on the \textit{Initial Task Plan} button to trigger the generation of personalized Weekly Task Schedule \textcircled{4}, along with task distribution and collaboration suggestions \textcircled{3}. ParPal first decomposes the overall task list into multiple detailed and actionable subtasks when the task allocation feature is triggered. It then assigns these subtasks to different caregivers and generates task-specific tutoring suggestions. These suggestions provide a summary of a collaborative learning plan and specify how caregivers can work together to complete the assigned tasks. 

This process consists of three main substeps:

\begin{itemize}
    \item \textbf{Substep 1: Task decomposition.} A structured prompt decomposes weekly learning goals into age-appropriate, actionable subtasks. Role prompting is used to define an expert persona in family education to improve task specificity and suitability~\cite{wang2023rolellm}.
    \item \textbf{Substep 2: Subtask scheduling.} Subtasks are assigned to caregivers and scheduled based on availability and expertise while preserving logical task order.
    \item \textbf{Substep 3: Conflict checking.} The system checks for time conflicts (e.g., overlapping slots) and produces a finalized schedule formatted in JSON and rendered as a shared weekly timesheet.
\end{itemize}

In addition to task allocation, ParPal generates a brief summary describing the weekly collaboration plan and task-specific tutoring suggestions.

\paragraph{Shared timesheet and progress tracking}

To address imbalanced division of labor in homework tutoring from formative study, a timesheet \textcircled{4} is designed to share task details. Meanwhile, when caregivers click on a subtask in the timesheet, a task card is displayed, showing the AI-generated task description, recommended tutoring method, task owner, expected start and end times, and task completion status \textcircled{5}. Due to variation in availability between working hours and personal time, ParPal enables caregivers to adjust or hand over subtasks generated by AI to another caregiver.

\subsubsection{Individual Task Support Panel Figure (\ref{Figure4})}

The Individual Task Support Panel provides AI-assisted tutoring support for completing assigned subtasks (\textbf{DS3}). For each task, caregivers receive an AI-generated task description and tutoring suggestions tailored to the child's grade level and learning context. Caregivers responsible for the same task can leave notes or record progress to support handover and coordination.

\paragraph{Interaction with tutoring assistance}

Caregivers can interact with a prompt-based chatbot powered by GPT-4o to ask questions, request clarification, or seek additional learning materials during the tutoring process. Additionally, ParPal provides three AI-driven features to reduce the burden of tutoring and address the' knowledge gaps of caregivers:

\textit{Answer checking.} Caregivers can upload photos of homework to verify answers without re-solving problems.
\textit{New example generation.} The system generates additional practice problems that share underlying concepts with the original task.
\textit{Tutoring guidance.} ParPal generates age-appropriate explanations and structured guidance to support caregivers in explaining problems effectively.

While answer checking and example generation resemble existing homework assistance platforms (e.g., Zuoyebang~\cite{zuoyebang2025}, Xiaoyuan Souti~\cite{xiaoyuandouti2025}), ParPal emphasizes \textit{caregiver-in-the-loop tutoring} by generating guidance that supports how caregivers explain problems rather than directly providing final answers.

\begin{figure*}[t]
  \centering
  \includegraphics[scale=0.40]{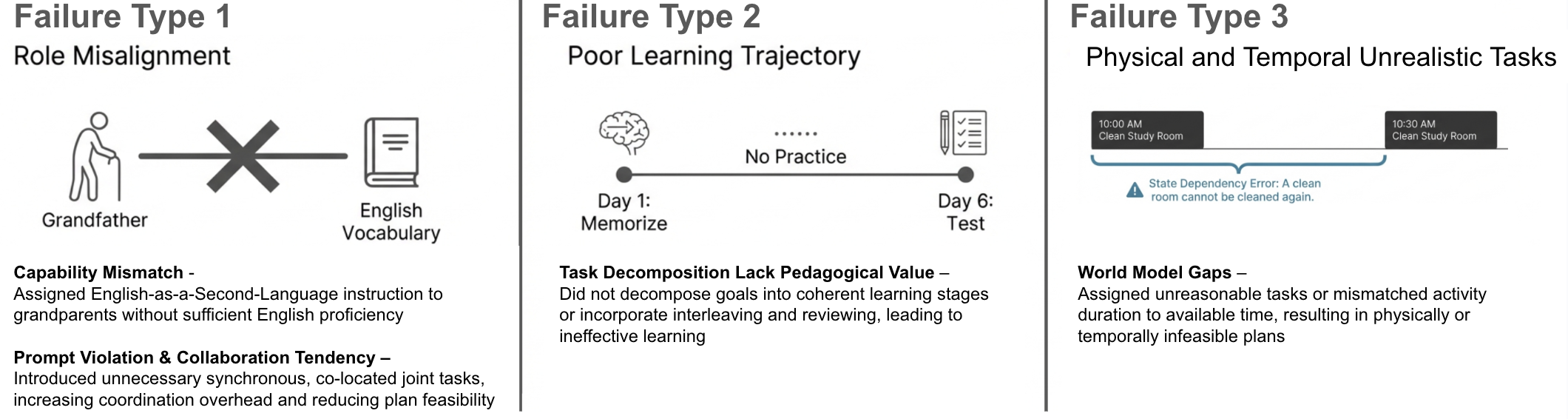}
  \caption{Common Reasons for Low Scoring Cases in task decomposition and allocation.}
  \label{Figure3}
\end{figure*}

\subsection{Expert Evaluations of Task Decomposition and Allocation Module (Core Feature)} Most existing evaluations of LLMs in education focus on specifci teaching and learning processes such as math problem solving, factual question answering, or one-to-one tutoring. We instead introduce learning plan generation as a novel task that requires coordinating multiple actors, reasoning about multi-subject learning activities and patterns, and respecting realistic execution constraints. While our task below does not aim to establish a benchmark, it provides an prelimary examination of a LLM's distinct capability and its limitations that are not captured by prior studies and motivates further systematic study.

\subsubsection{Evaluation Task Design} 

We designed \textbf{10 family profiles} to evaluate large language models' ability to generate
family-oriented weekly learning plans under realistic constraints. Each families composition specifies a
fixed \textbf{child profile} (age, grade level, and learning characteristics) and a fixed \textbf{family configuration}
(2--4 caregivers drawn from parents and grandparents). For each caregiver, we additionally provided
\textbf{explicit availability windows} (weekday vs.\ weekend, hour-level granularity) to capture practical
coordination constraints in home learning.

For each family composition, we evaluated \textbf{three sets of learning tasks} to combines a wide range
of at-home learning needs. These task sets were informed by our initial formative data collection
with families on what they expected children to accomplish at home (e.g., academic consolidation,
practice routines, habit formation, and reflective learning). Combining 10 family compositions with
five learning-task sets per composition, we obtained a total of 50 weekly learning plans per model. To enable a reasonable comparison, we used the same prompt and identical inputs when querying
GPT-4o and Claude 4.5.  The system was developed in July 2025, before GPT-5 was available.

\subsubsection{Expert Evaluation Metrics}

Two domain experts independently rated each generated weekly plan along five dimensions:
(i) \textbf{Role--Task Alignment} (who does what),
(ii) \textbf{Task Decomposition Quality} (how subtasks are structured across the day and week),
(iii) \textbf{Task Coverage / Completeness} (whether essential components are missing),
(iv) \textbf{Context Awareness} (whether the model correctly interpreted contextual constraints), and
(v) \textbf{Actionability} (whether the plan is feasible in practice).
Each dimension was rated on a 3-point ordinal scale and results are shown in Table \ref{tab:expert_ratings_compact}. 
In addition to numeric scores, the expert provided qualitative annotations explaining low ratings. Afterwards, the two experts jointly reviewed and discussed cases with discrepant scores to reach consensus.

We summarize the quantitative comparison below and present a qualitative synthesis of model-specific strengths and failure modes in the next section. Based on the combined results, GPT-4o was selected for subsequent system development, as it demonstrated stronger capability in distributing tasks across families with more than three caregivers, aligning with our system's goal of supporting effective communication and collaboration. Claude tended to generate more explicitly collaborative activities among family members (e.g., “summarize the previous week together”), which reflects good intentions but did not align with our requirement for independent tasks for different members. However, Claude performed better in task decomposition quality, often breaking a high-level learning task into concrete daily activities. Task Coverage / Completeness was overall great for both models. Action ability was relatively poor for both models. Overall, the expert did not observe significant differences between the two models.

\begin{table}[t]
\centering
\small
\setlength{\tabcolsep}{3pt}
\resizebox{\columnwidth}{!}{
\begin{tabular}{lcccc}
\hline
\textbf{Metric} 
& \textbf{4o Avg.} 
& \textbf{Claude Avg.} 
& \textbf{4o Best (\#)} 
& \textbf{Claude Best (\#)} \\
\hline
Role--Task Alignment 
& 2.87 & 2.87 & 3 (21) & 3 (21) \\

Task Decomposition 
& 2.67 & 2.80 & 3 (17) & 3 (17) \\

Task Coverage 
& 2.96 & 2.93 & 3 (21) & 3 (20) \\

Context Awareness 
& 2.83 & 2.87 & 3 (19) & 3 (20) \\

Actionability 
& 2.43 & 2.50 & 3 (10) & 3 (11) \\
\hline
\end{tabular}
}
\caption{Expert ratings of weekly learning plans across \textbf{10 family compositions} and \textbf{three learning task sets per composition} (\textbf{30 plans per model}). Ratings use a 3-point scale: 1 = Not acceptable, 2 = Needs improvement, 3 = Good. }
\label{tab:expert_ratings_compact}
\end{table}


\subsubsection{Case Studies from Evaluation} The following case studies focus on identifying why certain plans received lower scores, rather than comparing how often specific issues occurred across models. We highlight representative failure patterns and edge cases observed in the generated plans.

\paragraph{Role–Task Alignment: Misaligned Role Expertise and Unnecessary Collaborative Activities}
Low scores often occurred when generated plans failed to respect role-specific capabilities or the prompt requirement that tasks should remain independent. In one case, the input described the grandfather's role as \emph{monitoring and checking task completion} in a \textit{Chinese second-tier city} (translated from Chinese), yet the plan assigned him to mentor English vocabulary learning (e.g., ``15:00--16:00: grandfather leads English word learning,'' translated from Chinese), despite no indication of language proficiency. In edge cases, the module would still assign tasks to times where none of the caregivers would be available. 

In most cases, lower alignment stemmed from the model's limited understanding of collaboration (and how it could be designed to benefits learning). In one case, the model assigned tasks that implicitly required collaboration, such as scheduling a joint exercise session led by the father followed by a shared review and recording activity with the mother and child. While these activities were reasonable individually, they introduced cross-task dependencies, violating the independence constraint specified in the task prompt. Another example, for a 300-word reflective writing task, one plan assigned a joint revision session in which the mother focused on emotional expression while the father checked language and logic. Although collaboration can be useful, this division of labor was not warranted by the task and introduced unnecessary coordination. Similarly, another plan required both parents to jointly rate the child's Chinese and English performance and conduct an oral reflection. For a simple evaluative activity, simultaneous involvement of multiple caregivers does not seem common, collaborative and potentially improving learning outcomes and relationship. 

\paragraph{Task Decomposition: Lack of Learning Trajectory and Pedagogical Best Practices.}  Low task decomposition scores often arose when learning activities did not adequately supporting the underlying learning task and pedagogical best practices such as interleaving and revisiting content. For example, a learning task such as \emph{memorizing one English passage (8--10 sentences)} was decomposed into two sessions scheduled five days apart, without intermediate reinforcement, overlooking repetitions needed in second-language learning. Below is another example on second-language learning.
In the weaker plan, an \emph{English article memorization task intended to practice oral pronunciation} was decomposed primarily into: pronunciation-focused activities (Day 1), repeated full-text reading and recording comparison (Day 3), and isolated pronunciation drills (Day 5). In contrast, the stronger plan decomposed the same task into a staged sequence that explicitly supported memorization in everyday within the week from Day 1 to 7, beginning with focused work on the first half of the text, followed by completion of the second half, then repeated listening, comparison, and review. This structure better aligned pronunciation practice with memory development, leading to higher task decomposition quality scores.

Low scores also occurred when generated tasks made it unclear how the child was expected to participate or why certain caregivers were required. For example, given an input objective of \emph{memorizing one English song and learning 10 high-frequency words}, one plan scheduled a joint parent-led session in which the song was played, the mother explained the content, and the father annotated pronunciation points. However, the plan did not specify concrete actions for the child, such as singing, repeating lyrics, or practicing target words.

\paragraph{Actionability: Physically and Temperally Infeasible Tasks.}
Low actionability scores occurred when generated plans included tasks that were difficult or impossible to execute as specified. For example, one plan scheduled \emph{cleaning the study desk for the second and third time in the same 30 minute session}. Since desk organization is a state-based task that cannot be meaningfully repeated consecutively, assigning multiple consecutive rounds of the same physical activity was impractical.  In another case, a music rhythm practice activity was allocated an unrealistically short duration (e.g., ``\textit{20:00--20:05: music rhythm practice,'' }). By contrast, a more plausible plan allocated a longer continuous session (e.g., ``\textit{19:00--19:50: English children's song learning with both parents, including listening to the song, becoming familiar with the melody and lyrics, and explaining the content},'' , which better matched the time demands of the activity.

\section{End-User Evaluation with 11 Families}

\subsection{Participants and Recruitment}
We recruited 11 families through prior contacts and snowball sampling. In total, 23 individuals participated, including 11 mothers, 10 fathers, and 2 grandmothers. Participants were required to have prior experience tutoring children and to have at least one child attending elementary or junior high school. Each household received approximately \$30 USD as compensation. This study was approved by the institutional review board (IRB). Although our tool was designed to support more than two caregivers, we faced challenges recruiting families with three or more active caregivers willing to try out tool. In practice, grandparents were less likely to engage directly with the AI system and therefore did not participate in tool use. In some families, mothers instead delegated certain of their own tasks to grandparents through other channels, such as face-to-face communication.

\subsection{Study Procedure}


\subsubsection{Field study}
Each family used \tool{} for one week, with six families voluntarily extending participation to two weeks. We checked in periodically to address questions or issues and encouraged participants to integrate the system into their everyday tutoring activities. Individual family's usage of different features are shown in Table \ref{usage}. Overall, the Task Distribution was useful for some families but not all of them. 

We conducted post-study interviews with all families, each lasting approximately 60 minutes. Interviews focused on caregivers' experiences using \tool{}, its role in supporting collaboration and tutoring practices, and its perceived impact on caregiving labor, coordination, emotional support, and children's learning outcomes.

All interviews were conducted in Mandarin and audio-recorded with consent. We analyzed the transcripts using thematic analysis~\cite{braun2006using}. Two authors independently coded a subset of interviews and iteratively refined the codebook through discussion until consensus was reached.

\subsection{Results}
Analysis of the system log and interview data revealed several key takeaways about how our tools can support family learning:

\textbf{Making hidden caregiving labor visible.} Caregivers described how \tool{} helped surface often invisible planning and coordination work, allowing family members to better recognize each other's contributions.

\textbf{Supporting coordination among caregivers.} The system facilitated shared understanding of tutoring responsibilities, reducing ambiguity and conflict when multiple caregivers were involved.

\textbf{Aligning diverse family perspectives.} \tool{} served as a mediating artifact that helped caregivers negotiate differing expectations, priorities, and tutoring strategies.

\textbf{Providing emotional reassurance.} Caregivers reported that interacting with \tool{} offered emotional support by validating their concerns and reducing anxiety around tutoring decisions.

\textbf{Shaping tutoring strategies over time.} Through continued use, caregivers reflected on and adjusted their tutoring approaches, leading to more deliberate and coordinated family learning practices.

These findings show how human-centered AI systems can function not only as instructional aids, but also as supports that shape family coordination, emotional experiences, and caregiving practices around children's learning.

\subsubsection{ Suggestions for Improving Task Decomposition and Allocation}
\begin{table}[t]
\centering
\footnotesize
\setlength{\tabcolsep}{5pt}
\renewcommand{\arraystretch}{1.0}
\caption{Caregiver engagement with task distribution and individual task support ($N=23$ caregivers).}
\label{tab:engagement-summary}
\begin{tabular}{lcc}
\toprule
\textbf{Caregiver Engagement} & \textbf{} & \textbf{} \\
\midrule
\multicolumn{3}{l}{\textit{Task Distribution — engagement intensity}} \\
Tasks completed        & 1 / 1 / 6   & Min / Med / Max \\
Subtasks executed      & 1 / 4 / 31  & Min / Med / Max \\
\midrule
\multicolumn{3}{l}{\textit{Individual Task Support — caregiver coverage}} \\
New example            & 50\%        & Caregivers \\
Answer checking        & 45\%        & Caregivers \\
Tutoring guidance      & 50\%        & Caregivers \\
\bottomrule
\end{tabular}

\vspace{2pt}
\raggedright
\scriptsize
\end{table}

While our findings highlight the positivities of AI-supported family learning, they also reveal several challenges that point to important opportunities for improving the design of AI-mediated family learning tools. Specifically, in Table \ref{usage}, over half of the families continue to use "Individual Task Support" but no longer execute the sub-tasks decomposed and allocated by ``Task Distribution Panel''. In particular, participants expressed dissatisfaction with the system's task assignment mechanisms as described in the previous section. 

\paragraph{Dynamic Contexts Call for Incremental, Context-Aware Re-Planning}
Caregiving contexts vary in caregiver expertise, time availability, and stress level, and these factors can change even minute and planning ahead may be useful but less actionable. As a result, static support strategies are often ineffective: assistance that works during planned study time may become disruptive when time is limited. While \tool{} helped shift attention from task completion to conceptual understanding, participants wanted more precise control over the amount, form, and timing of assistance. Algorithmically, this suggests the need for context-aware planning that supports incremental re-planning, where parts of a task schedule are updated based on observed signals (e.g., delays, skipped steps, or user overrides) rather than recomputed from scratch. Previous contribution histories of family members can be included in the planning state, allowing the system to adjust plans in more targeted ways, such as breaking tasks into smaller units or suggesting short, immediately actionable activities that fit current constraints.

\paragraph{Longitudinal Modeling and Mediation of Collaborative Effort in Multi-Actor Contexts. } Participants observed that measuring effort solely through per-user task completion produced incomplete and sometimes misleading estimates of contribution. The system primarily logged explicit actions (e.g., completing assigned tasks) and offered limited visibility into coordination behaviors such as task delegation, handoffs, or last-minute reassignment. As a result, per-user metrics failed to capture how effort was redistributed over time or shared across family members, and dashboard-style summaries were sometimes interpreted as competitive or evaluative feedback. These observations suggest modeling collaborative effort as a group-level construct inferred from task structure and temporal interaction patterns, rather than as a simple count of completed tasks. Accordingly, the system should maintain longitudinal profiles that capture coordination-related behaviors alongside task outcomes. Such representations can support adaptive task redistribution—for example, assigning smaller or complementary tasks to less-engaged members—while accounting for invisible coordination work and encouraging engagement without relying on explicitly competitive metrics or evaluative dashboards.
\section{Discussions and Implications}
\subsection{Multi-Actor Plan Generation as a New Problem Space  for LLMs}

Family learning exposes planning challenges that are largely absent from existing LLM benchmarks and AI-in-education evaluations. Unlike classroom or single-learner settings, it involves multiple actors with asymmetric expertise, informal instructional decision-making, and soft, evolving constraints such as availability, fatigue, and competing responsibilities. Learning goals are negotiated rather than fixed, and plans must remain usable under partial execution and frequent disruption. Our findings show that current commercial zero-shot LLMs struggle in this setting, often failing to produce plans that are simultaneously role-appropriate, well-decomposed, and temporally or physically feasible. These limitations are not well captured by benchmarks centered on static goals or single-user task completion. We argue that multi-actor family learning provides a critical benchmark for evaluating LLM-based planning under realistic human and contextual constraints.

\subsection{Design Implication: Modeling of Collaboration and Social Cost}

Our findings indicate that current LLM-based planners fail not because they lack pedagogical knowledge, but because they lack a representation of coordination overhead. When multiple caregivers are present, models violate prompts to create individual tasks and routinely introduce joint or synchronized activities without accounting for the social cost of coordination—such as scheduling constraints, effort imbalance, and cognitive load. This suggests a structural limitation in current multi-actor planning architectures: collaboration is treated as implicitly beneficial rather than as a costly resource that must be justified. Future human–AI planning systems should therefore incorporate an explicit social cost function that enables models to reason about when collaboration is necessary, when independent action is preferable, and how coordination trade-offs affect plan feasibility and robustness.

\subsection{LLMs Have Limited Pedagogical Knowledge}Our expert evaluation shows that LLMs can describe good teaching but struggle to plan it. Generated plans often include reasonable activities yet fail to form coherent learning progressions over time, missing sequencing, reinforcement, interleaving, or revisiting. Stronger plans stood out not by better wording, but by better temporal structure: they staged activities across days, aligned them with clear goals, and used repetition intentionally. This pattern suggests that pedagogical knowledge in LLMs remains largely linguistic rather than procedural. 

Additionally, low actionability reflects a separate breakdown in real-world reasoning. Models frequently misjudge time and effort, produce unrealistic schedules, and assume ideal execution without mechanisms for adjustment. Together, these limitations point to a gap between instructional knowledge and actionable planning. Addressing this gap requires AI systems that reason explicitly about temporal structure, everyday constraints, and plan adaptation—core requirements for human-centered AI in real learning settings.
\section{Conclusions}
This study investigates how LLM-powered learning systems can support family learning - a setting characterized by multiple actors, different expertise and shifting availability. Through a formative study and a field study with 11 families demonstrates that such AI support can improve coordination clarity and shared responsibility in everyday learning, while also exposing limitations of current AI systems in modeling pedagogical structure, real-world feasibility, and collaborative effort. Together, these findings suggest that effective AI for education must move beyond child/student-centric tutoring to account for collaboration, planning, and caregiving practices.

\bibliographystyle{named}
\bibliography{ijcai26}

\appendix
\onecolumn

\section{Additional System User Interfaces}

\begin{figure}[t]
  \centering
  \includegraphics[scale=0.15]{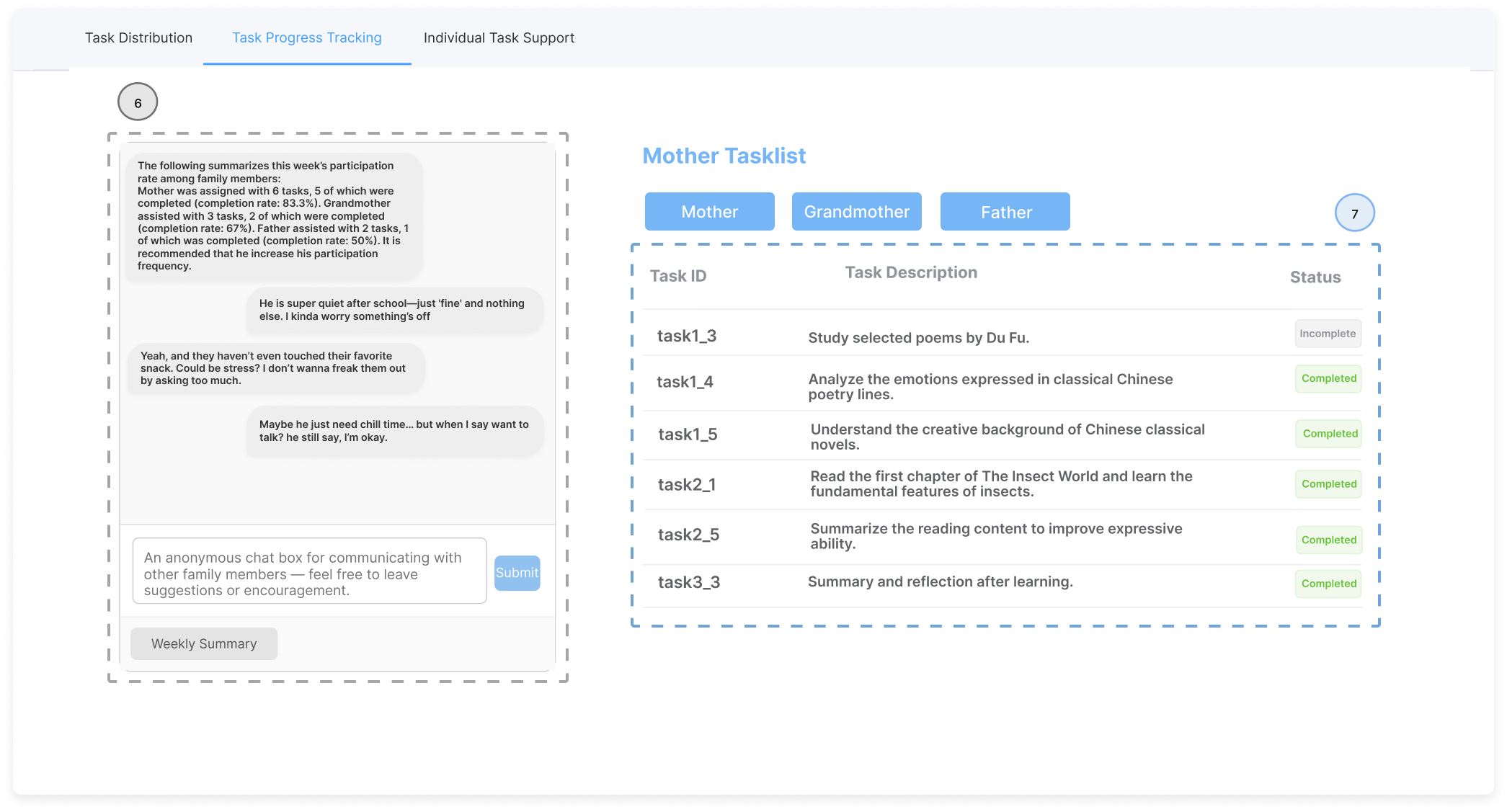}
  \caption{ParPal's dashboard to support task progress tracking and summary.}
  \label{Figure3}
\end{figure}

\begin{figure}[t]
  \centering
  \includegraphics[scale=0.15]{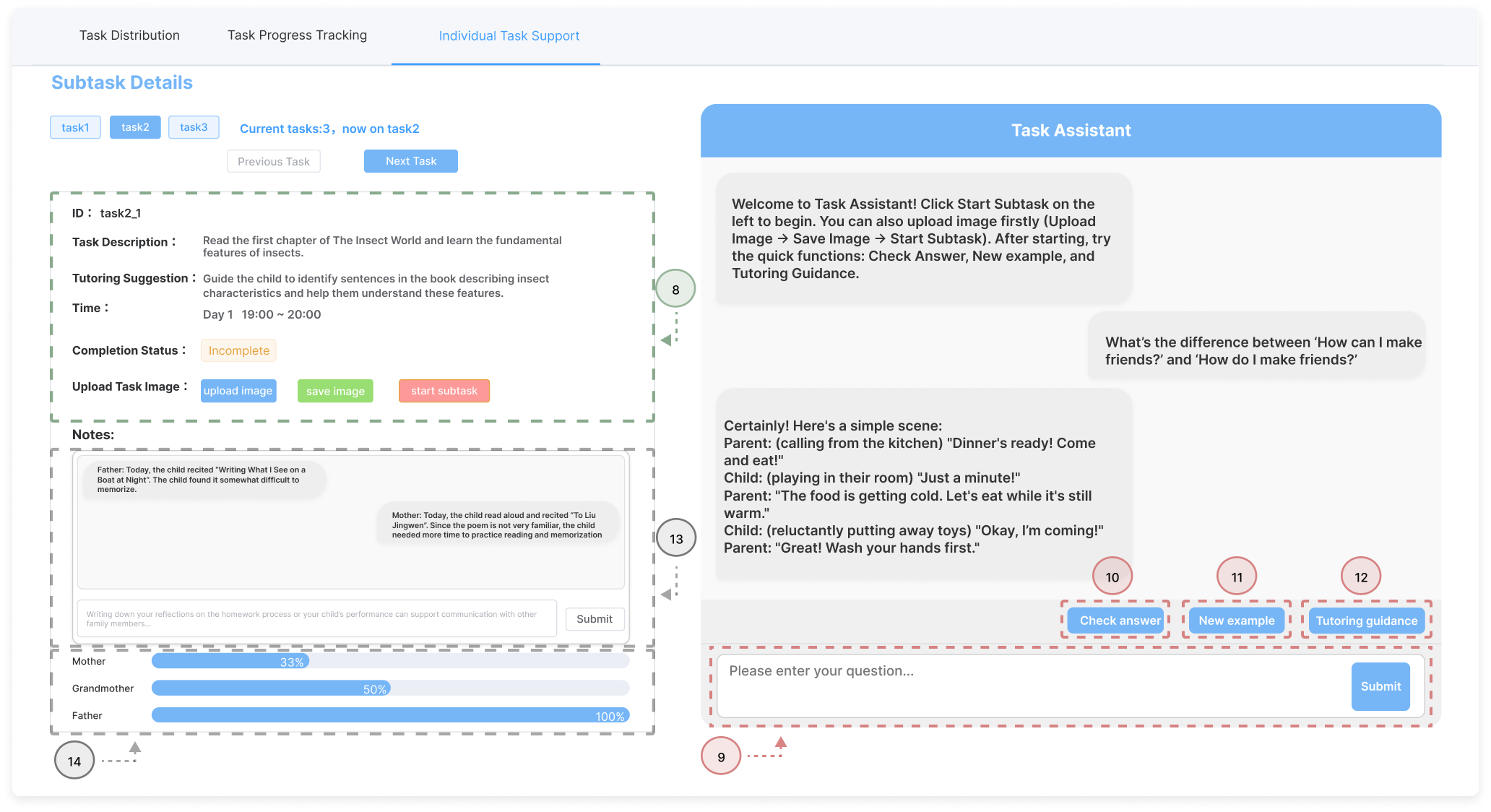}
  \caption{ParPal's Individual Task Support Panel to support tutoring guidance.}\
  \label{Figure4}
\end{figure}

\section{Prompts}

Translated from Chinese (Original) to English

\subsubsection{Stage 1: Task Decomposition and Scheduling}

Stage~1 consists of a task decomposition phase and a planning phase.
Four prompts are used in total, each invoking the OpenAI API once.
Together, they generate a structured task schedule and a final
high-level collaboration suggestion.

\paragraph{Prompt 1: Task Decomposition (build\_task\_split)}

This prompt decomposes an initial family task into multiple subtasks
suitable for collaborative completion by parents and a child.

\begin{lstlisting}
You are an expert in elementary-school family homework planning.
Your task is to help parents plan homework collaboratively.

Decompose the following task:
"{task_name}: {task_description}"

into multiple subtasks suitable for parents and a child to complete
together. For each subtask, answer all questions involved.

Requirements:
- Each subtask must have at least one responsible adult.
- Task assignment must follow family preferences: {family_desc}.
- Consider the child's cognitive level and local context: {child_desc}.
- Group similar questions; total subtasks must not exceed 10.
- Provide task description, answers, and tutoring strategies.
- Every family member must participate.
- Subtask names must follow:
  {task_name}_1, {task_name}_2, ...

Return valid JSON following:
{parser.get_format_instructions()}
\end{lstlisting}

\paragraph{Prompt 2: Time Allocation for Subtasks}

This prompt assigns each subtask to a specific day and time.

\begin{lstlisting}
Schedule each subtask with a specific day and time.

If the child can complete a task independently, it may be scheduled
during daytime hours without considering adult availability.

Family member availability:
{members}

Scheduling rules:
- Each task must be assigned to exactly one day (Day 1--Day 7).
- Every day must include at least one task.
- Each task must include start and end times (HH:MM).
- Do not modify task content; only add Day, start, and end fields.
- Tasks should be distributed across morning, afternoon, and evening.
- Consider task order and dependencies.

Return valid JSON following:
{format_instructions}

Task list:
{task_assignment_dict}
\end{lstlisting}

\paragraph{Prompt 3: Conflict Detection and Rescheduling}

This prompt checks for time conflicts and resolves them.

\begin{lstlisting}
Check all subtasks for scheduling conflicts, including:
- Overlapping time intervals
- Multiple tasks at the same time
- Responsible members being unavailable

Instructions:
- Only modify Day, start time, and end time.
- Do not change task descriptions or responsibilities.
- No overlapping tasks are allowed.
- Use 24-hour time format (HH:MM).
- Tasks must be scheduled within Day 1--Day 7.
- Every day must include at least one task.

Return an updated valid JSON structure.

Initial schedule:
{schedule_dict_all}

Family availability:
{members}
\end{lstlisting}

\paragraph{Prompt 4: General Collaboration Suggestion}

This prompt generates a high-level summary describing how the family
should collaborate across all subtasks.

\begin{lstlisting}
You are a family collaboration planning advisor.

Generate a general suggestion that:
- Summarizes the subtasks involved
- Describes how family members collaborate
- Uses warm and supportive language

Output a single paragraph in Chinese (no JSON),
limited to 300 characters.

Final schedule:
{final_schedule}

Family members:
{members}
\end{lstlisting}

\subsubsection{Stage 2: Parent--Chatbot Interaction}

Stage~2 enables direct interaction between parents and the chatbot.
It includes a general dialogue mode and three specialized assistance modes.

\paragraph{Prompt 5: General Dialogue Mode}

\begin{lstlisting}
You are an experienced homework tutoring assistant.
Help parents complete the homework and answer questions clearly
and concisely. Limit responses to 100 words.
\end{lstlisting}

\paragraph{Prompt 6: Homework Checking Mode}

\begin{lstlisting}
Below is the child's homework.
Check whether the answers are correct and provide improvement
suggestions for parents.
\end{lstlisting}

\paragraph{Prompt 7: Transfer Practice Generation}

\begin{lstlisting}
Below are the child's completed answers.
For each question, generate 1--2 similar practice problems.
Use the same question type and knowledge point, suitable for
elementary students.
\end{lstlisting}

\paragraph{Prompt 8: Parent Explanation Support}

\begin{lstlisting}
Provide guidance for parents on how to explain the following
problems to an elementary student.
Use simple and age-appropriate language.
\end{lstlisting}

\section{11 Family Information}

\begin{table}[]
\small
\caption{Participants' demographics.}
\label{Demongraphic information} 
\resizebox{\columnwidth}{!}{%
\begin{tabular}{p{1cm}p{1.5cm}p{1cm}p{1cm}p{2cm}p{2cm}p{3.2cm}p{0.5cm}p{0.6cm}p{0.8cm}p{0.5cm}}
\toprule
\textbf{Family ID} & \textbf{\shortstack{Caregiver ID$^{*}$}} & \textbf{Role} & \textbf{Age Group} & \textbf{Occupation}  & \textbf{Education Level} & \textbf{AI Learning Tools Used}& \textbf{Child ID} & \textbf{Grade} & \textbf{Gender} & \textbf{Age}

\\ \hline

\multirow{3}{*}{Family 1} & F1D & Father & 41-45 & Goverment Job & Undergraduate & \multirow{3}{*}{Zuoyebang, Doubao, Baidu AI} & \multirow{3}{*}{FK1} & \multirow{3}{*}{4} & \multirow{3}{*}{Female} & \multirow{3}{*}{10} \\ \cline{2-6}
& F1M & Mother & 36-40 & Goverment Job & Undergraduate &  &  &  &  &  \\ \cline{2-6}
& F1GM & Grandma & 61-65 & Retired & Not Disclosed &  &  &  &  &  \\ \hline
\multirow{2}{*}{Family 2} & \shortstack{F2M$^{*}$} & Mother & 36-40 & Teacher & Undergraduate  & \multirow{2}{*}{Zuoyebang} & \multirow{2}{*}{FK2} & \multirow{2}{*}{5} & \multirow{2}{*}{Female} & \multirow{2}{*}{11} \\ \cline{2-6}
& F2D & Father & 36-40 & Technical Staff & Undergraduate&  &  &  &  &  \\ \hline
\multirow{2}{*}{Family 3} & \shortstack{F3M$^{*}$} & Mother & 41-45 & Cost Consulting & Undergraduate & \multirow{2}{*}{Zuoyebang, Yuanfudao} & \multirow{2}{*}{FK3} & \multirow{2}{*}{4} & \multirow{2}{*}{Male} & \multirow{2}{*}{9} \\ \cline{2-6}
& F3D & Father & 41-45 & Landscaping & Junior College &  &  &  &  &  \\ \hline
\multirow{2}{*}{Family 4}& \shortstack{F4M$^{*}$} & Mother & 41-45 & Teacher & Undergraduate & \multirow{2}{*}{Zuoyebang, Yuanfudao} & \multirow{2}{*}{FK4} & \multirow{2}{*}{6} & \multirow{2}{*}{Male} & \multirow{2}{*}{12} \\ \cline{2-6}
& F4GM & Grandma & 70-74 & Retired & High School&  &  &  &  &  \\ \hline
\multirow{2}{*}{Family 5} & \shortstack{F5M$^{*}$} & Mother & 36-40 & Professor & PhD & \multirow{2}{*}{ChatGPT} & \multirow{2}{*}{FK5} & \multirow{2}{*}{3} & \multirow{2}{*}{Male} & \multirow{2}{*}{8} \\ \cline{2-6}
& F5D & Father & 36-40 & Engineer & PhD &   &  &  &  &  \\ \hline
\multirow{2}{*}{Family 6} & \shortstack{F6M$^{*}$} & Mother & 36-40 & Staff & Undergraduate & \multirow{2}{*}{Zuoyebang} & \multirow{2}{*}{FK6} & \multirow{2}{*}{5} & \multirow{2}{*}{Female} & \multirow{2}{*}{11} \\ \cline{2-6}
& F6D & Father & 36-40 & Staff & Undergraduate &  &  &  &  &  \\ \hline
\multirow{2}{*}{Family 7} & F7M & Mother & 41-45 & Teacher & Undergraduate & \multirow{2}{*}{Doubao, Ernie Bot} & \multirow{2}{*}{FK7} & \multirow{2}{*}{4} & \multirow{2}{*}{Female} & \multirow{2}{*}{10} \\ \cline{2-6}
& F7D & Father & 41-45 & Engineering & Undergraduate &  &  &  &  &  \\ \hline
\multirow{2}{*}{Family 8} & F8M & Mother & 36-40 & Teacher & Undergraduate & \multirow{2}{*}{Doubao} & \multirow{2}{*}{FK8} & \multirow{2}{*}{4} & \multirow{2}{*}{Male} & \multirow{2}{*}{10} \\ \cline{2-6}
& \shortstack{F8D$^{*}$} & Father & 36-40 & Engineering & Undergraduate &  &  &  &  \\ \hline
\multirow{2}{*}{Family 9} & \shortstack{F9M$^{*}$} & Mother & 36-40 & HR & Undergraduate & \multirow{2}{*}{Deepseek, Yuanfudao} & \multirow{2}{*}{FK9} & \multirow{2}{*}{6} & \multirow{2}{*}{Male} & \multirow{2}{*}{11} \\ \cline{2-6}
& F9D & Father & 41-45 & Lecturer & Graduate &  &  &  &  &  \\ \hline
\multirow{2}{*}{Family 10} & F10D & Father & 36-40 & Technical Staff & Undergraduate & \multirow{2}{*}{Zuoyebang, Yuanfudao} & \multirow{2}{*}{FK10} & \multirow{2}{*}{3} & \multirow{2}{*}{Male} & \multirow{2}{*}{9} \\ \cline{2-6}
& F10M & Mother & 36-40 & Telecommunications & Undergraduate &  &  &  &  &  \\ \hline
\multirow{2}{*}{Family 11} & F11D & Father & \multicolumn{1}{l}{41-45} & Security Officer & High School & \multirow{2}{*}{Baidu AI, Xiaoai classmate} & \multirow{2}{*}{FK11} & \multirow{2}{*}{5} & \multirow{2}{*}{Male} & \multirow{2}{*}{11} \\ \cline{2-6}
& \shortstack{F11M$^{*}$} & Mother & \multicolumn{1}{l}{36-40} & Teacher & Undergraduate &  &  &  &  &  \\ \bottomrule
\end{tabular}
}
{\raggedright \small* Primary caregiver, who takes the leading role in daily family learning. Without *, do not have a clearly identified primary caregiver.}
\end{table}

\begin{table}[t]
\centering
\tiny
\setlength{\tabcolsep}{3pt}
\renewcommand{\arraystretch}{1.15}
\caption{Task distribution and tutoring assistance across families and participants.
\textbf{Task Distribution} reports the number of completed tasks and executed subtasks.
\textbf{Individual Task Support} records whether specific tutoring strategies were applied.}
\label{tab:family-tutoring}
\begin{tabular}{@{}cc cc ccc@{}}
\toprule
\multirow{2}{*}{\makecell{\textbf{Family}\\\textbf{ID}}} &
\multirow{2}{*}{\makecell{\textbf{Caregiver}\\\textbf{ID}}} &
\multicolumn{2}{c}{\textbf{Task Distribution}} &
\multicolumn{3}{c}{\textbf{Individual Task Support}} \\
\cmidrule(lr){3-4}\cmidrule(lr){5-7}
& &
\makecell{\textbf{Tasks}\\\textbf{Completed}\\(\#)} &
\makecell{\textbf{Subtasks}\\\textbf{Executed}\\(\#)} &
\makecell{\textbf{New}\\\textbf{Example}} &
\makecell{\textbf{Answer}\\\textbf{Checking}} &
\makecell{\textbf{Tutoring}\\\textbf{Guidance}} \\
\midrule

\multirow{3}{*}{F1} & F1D  & \multirow{3}{*}{6} & 6  & Y & Y & Y \\
                   & F1M  &                      & 10 & Y & Y & Y \\
                   & F1GM &                      & 19 & Y & Y & Y \\
\midrule

\multirow{2}{*}{F2} & F2D  & \multirow{2}{*}{6} & 23 & Y & Y & Y \\
                   & F2M  &                      & 31 & Y & Y & Y \\
\midrule

\multirow{2}{*}{F3} & F3D  & \multirow{2}{*}{4} & 11 & Y & Y & Y \\
                   & F3M  &                      & 7  & Y & Y & Y \\
\midrule

\multirow{2}{*}{F4} & F4GM & \multirow{2}{*}{3} & 5  & Y & Y & Y \\
                   & F4M  &                      & 4  & N & N & Y \\
\midrule

\multirow{2}{*}{F5} & F5D  & \multirow{2}{*}{1} & 2  & N & N & N \\
                   & F5M  &                      & 5  & Y & N & Y \\
\midrule

\multirow{2}{*}{F6} & F6D  & \multirow{2}{*}{4} & 4  & N & N & N \\
                   & F6M  &                      & 14 & Y & Y & Y \\
\midrule

\multirow{2}{*}{F7} & F7D  & \multirow{2}{*}{1} & 1  & N & N & Y \\
                   & F7M  &                      & 1  & N & N & N \\
\midrule

\multirow{2}{*}{F8} & F8D  & \multirow{2}{*}{1} & 2  & N & N & N \\
                   & F8M  &                      & 2  & N & Y & Y \\
\midrule

\multirow{2}{*}{F9} & F9D  & \multirow{2}{*}{1} & 1  & N & Y & N \\
                   & F9M  &                      & 2  & N & Y & N \\
\midrule

\multirow{2}{*}{F10} & F10D & \multirow{2}{*}{1} & 1  & N & Y & N \\
                    & F10M &                      & 2  & N & Y & N \\
\midrule

\multirow{2}{*}{F11} & F11D & \multirow{2}{*}{1} & 4  & N & N & N \\
                    & F11M &                      & 4  & Y & Y & N \\
\bottomrule
\end{tabular}
\label{usage}
\end{table}

\end{document}